\documentstyle[a4,epsf,12pt]{article}
\newcommand{\beq}{\begin{equation}}
\newcommand{\eeqn}{\end{eqnarray}}
\newcommand{\beqn}{\begin{eqnarray}}
\newcommand{\eeq}{\end{equation}}
\newcommand{\np}{Nucl. Phys. \underline}
\newcommand{\pl}{Phys. Lett. \underline}
\newcommand{\mev}{{\rm MeV}}
\newcommand{\gev}{{\rm GeV}}
\newcommand{\wilson}{{\rm Wilson\ Action}}
\newcommand{\dl}{\stackrel{\leftarrow}{D}}
\newcommand{\dr}{\stackrel{\rightarrow}{D}}
\newcommand{\clover}{{\rm Clover\ Action}}
\newcommand{\err}[2]{{\footnotesize {$\;\begin{array}{@{}l@{}}
			  +\makebox[1.3em][r]{#1} \\[-0.6em]
			  -\makebox[1.3em][r]{#2}
			\end{array}$}}}
\begin{document}
\begin{titlepage}

\begin{flushright}
Southampton Preprint: SHEP 91/92-27\\
Edinburgh Preprint: 92/510\\
\end{flushright}
\vspace{5mm}
\begin{center}
{\Huge The Hyperfine Splitting in Charmonium:\\
Lattice Computations using the Wilson and Clover Fermion Actions}\\[15mm]
{\large\it UKQCD Collaboration}\\[3mm]

{\bf C.R.~Allton, C.T.~Sachrajda}\\
Physics Department, The University, Southampton SO9~5NH, UK

{\bf S.P.~Booth, K.C.~Bowler, D.S.~Henty,
R.D.~Kenway, B.J.~Pendleton, D.G.~Richards, J.N.~Simone,
A.D.~Simpson}\\
Department of Physics, The University of Edinburgh, Edinburgh EH9~3JZ,
Scotland

{\bf C.~Michael, P.W.~Stephenson}\\
DAMTP, University of Liverpool, Liverpool L69~3BX, UK

\end{center}
\vspace{5mm}

\begin{abstract}
We compute the hyperfine splitting $m_{J/\psi}-m_{\eta_c}$ on the
lattice, using both the Wilson and $O(a)$-improved (clover)
actions for quenched quarks.
The computations are performed on a $24^3\times48$ lattice at
$\beta = 6.2$, using
the same set of 18 gluon configurations for both fermion actions.
We find that the splitting is 1.83\err{13}{15} times larger with
the clover action than with the Wilson action, demonstrating the
sensitivity of the spin-splitting to the magnetic moment term
which is present in the clover action.
However, even with the clover action the result is less than half
of the physical mass-splitting.
We also compute the decay constants $f_{\eta_c}$ and $f^{-1}_{J/\psi}$,
both of which are considerably larger when computed using the clover
action than with the Wilson action.
For example for the ratio $f^{-1}_{J/\psi}/f^{-1}_{\rho}$ we find
0.32\err{1}{2} with the Wilson action and $0.48\pm 3$ with the
clover action
(the physical value is 0.44(2)).
\end{abstract}

\end{titlepage}

\paragraph {The Hyperfine Splitting} Lattice computations of the
vector-pseudoscalar mass-splitt\-ings $m_{D^*}-m_D$ and
$m_{J/\psi}-m_{\eta_c}$, using the standard Wilson action for the
quarks in the quenched approximation, give results which are much too
small \cite{hqs,sixpt4}.  At $\beta=6.2$, for which the inverse
lattice spacing ($a^{-1}$) is approximately 2.7 GeV, the discrepancy
is about a factor of 2 for $m_{D^*}-m_D$ and a factor of about 4 for
$m_{J/\psi}-m_{\eta_c}$.  In this letter we compute the hyperfine
splitting $m_{J/\psi}- m_{\eta_c}$ using both the Wilson fermion
action,
\beqn
S_F^W  & =  & a^4\sum _x \frac{1}{a}\Biggl\{\bar{q}(x)q(x)
             +\kappa\sum _\mu\Bigl[
            \bar{q}(x)(\gamma _\mu -r)U_\mu (x) q(x+\hat\mu )
\nonumber\\
   & &\hspace{0.3in} -\ \bar{q}(x+\hat\mu )(\gamma _\mu + r)
   U^\dagger _\mu(x)
            q(x)\Bigr]\Biggr\}
\label{eq:sfw}\eeqn
and the nearest-neighbour $O(a)$--improved (or ``clover") fermion
action \cite{sw},
\begin{equation}
S_F^C  = S_F^W - irg_0\kappa\frac{a}{2}a^4\sum_{x,\mu,\nu}\bar{q}(x)
         F_{\mu\nu}(x)\sigma_{\mu\nu}q(x)
\label{eq:sclover}\end{equation}
with the same set of 18 gluon configurations in each case.  The
computations are performed on a $24^3\times 48$ lattice at $\beta=6.2$
with $r=1$.  These 18 configurations have been used earlier in our
study of light quark spectroscopy and meson decay constants, the
results and computational details can be found in
ref.\cite{ukqcd1,light_hadrons}.  For the quantities studied in
\cite{ukqcd1} the results obtained with the two actions were broadly
compatible.  However, here we show that the hyperfine splitting in
charmonium is almost a factor of 2 larger with the clover action than
with the Wilson action (see eq.(\ref{eq:ratio}) below), demonstrating
the sensitivity of this quantity to the magnetic moment term in
eq.(\ref{eq:sclover}).  We also present the results for the decay
constants of the $\eta_c$ and $J/\psi$ mesons.

Qualitatively similar results for the hyperfine splitting were
obtained by the Fermilab group \cite{aida}, who have performed
simulations using a fermion action which is similar to that in
equation~\ref{eq:sclover} but with a factor of 1.4 multiplying the
second term.  This factor is their estimate of the effects of higher
order perturbative corrections, and was obtained using a mean field
theory calculation \cite{ehkm}.  These authors also find a larger
value of the hyperfine splitting with their action than with the
Wilson action. However this comparison is obtained from computations
on lattices of different size.  Below we will compare our results with
those of ref.\cite{aida}.

For this study we took $\kappa = 0.1350$ for the Wilson action and
$\kappa = 0.1290$ for the clover action.  These values were chosen so
that the pseudoscalar meson masses are almost equal for both actions,
and correspond approximately to the physical mass of the $\eta_c$.  In
table \ref{tab:masses} we present our results for the masses of the
vector and pseudoscalar mesons (in lattice units), and for their
difference.  Setting the scale from the string tension gives
$a^{-1}=2.73(5)\,
\gev$ \cite{ukqcd1}
\footnote{The values of the inverse lattice spacing obtained by
comparing the lattice values of the masses of the light hadrons to the
physical ones lie within 15\% of the result from the string tension.
We take this as an indication of the size of the systematic
uncertainty.}, and using this value we see that the masses of the
mesons are within a few percent of their physical values.
\begin{table}
\centering
\begin{tabular} {|c|c|c|c|} \hline
 & $m_{\eta_c}a$ & $m_{J/\psi}a$ & $(m_{J/\psi}-m_{\eta_c})a$ \\
 \hline
Wilson & 1.066\err{6}{4} & 1.076\err{4}{5} & 0.0104\err{8}{7} \\
clover & 1.071\err{6}{4} & 1.088\err{5}{5} & 0.0190\err{12}{16} \\ \hline
\end{tabular}
\caption{Masses (in lattice units) of the pseudoscalar and vector
heavy-heavy mesons.}
\label{tab:masses}
\end{table}
The masses were obtained by fitting the correlation functions in
the time range $t=14-20$, and the mass differences were obtained
by fitting the ratio of the vector and pseudoscalar propagators
over the same range.
The fits were performed taking the correlations between the values
at different time slices into account, and the values of $\chi^2/$d.o.f
were acceptable.
The errors presented in this letter were obtained using the bootstrap
procedure, described in detail in ref.\cite{ukqcd1,light_hadrons}.

The main result of this letter comes from the entries in the
third column of table \ref{tab:masses}, from which it is clear that
the hyperfine splittings are very different for the two actions.
We stress that the results were obtained using the same gluon
configurations and with the same analysis techniques.
It is therefore likely that a number of the systematic errors
would cancel in the ratio, for which we find:
\beq
\frac{(m_{J/\psi}-m_{\eta_c})^{\rm clover}}
                              {(m_{J/\psi}-m_{\eta_c})^{\rm Wilson}}
                                                 =1.83\mbox{\err{13}{15}}
\label{eq:ratio}\eeq

\begin{figure}[htbp]
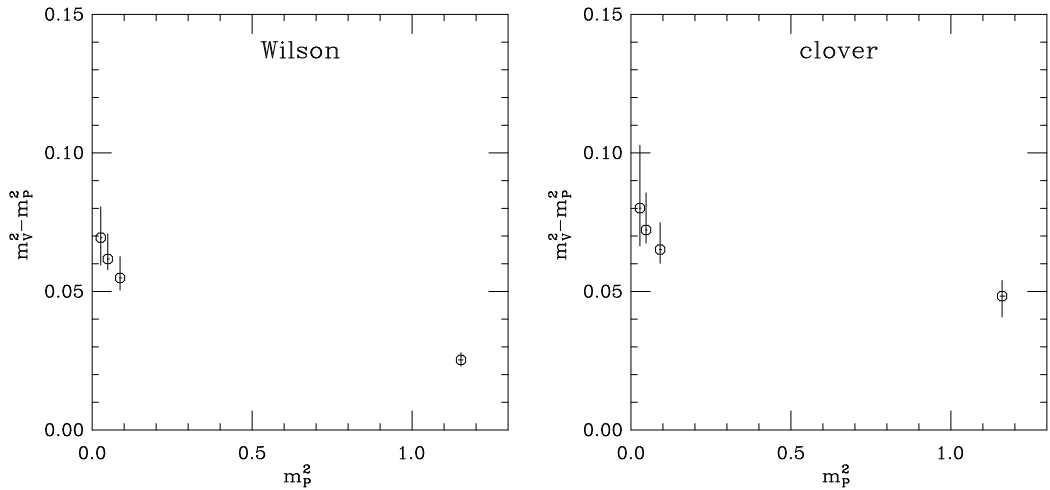

\centerline{
\setlength\epsfxsize{200pt}
  \epsfbox[20 30 620 600]{CHARM_mV2_minus_mP2_vs_mP2_WLL.ps}
\setlength\epsfxsize{200pt}
  \epsfbox[20 30 620 600]{CHARM_mV2_minus_mP2_vs_mP2_CLL.ps}
}
\caption{$m_V^2-m_P^2$ versus $m_P^2$, in lattice units, for the
Wilson and clover actions.\label{vec-pseudo}}
\end{figure}

In fig.~\ref{vec-pseudo} we plot the values of $m_V^2-m_P^2$ (where
$V$ and $P$ represent vector and pseudoscalar respectively) as a
function of the square of the pseudoscalar mass.  We include not only
the values for charmonium obtained from table~\ref{tab:masses}, but
also those for mesons composed of light quarks for three different
light quark masses \cite{ukqcd1}.  We see that for small masses the
two actions give similar results, but as the mass is increased a gap
gradually opens, with the clover action giving a larger value for the
hyperfine splitting.
\epsfverbosetrue

It is interesting to note that the suggestion that this quantity might
be susceptible to lattice artefacts has been made previously by the
APE collaboration in a comparative study~\cite{APE} of Wilson and
staggered fermion actions, albeit at stronger coupling and lighter
quark masses than the present work.

The result for the hyperfine splitting of charmonium obtained
with the clover action is still much smaller than the experimental
value.
Taking $a^{-1}=2.73\,\gev$, the values in table \ref{tab:masses}
correspond to:
\beqn
m_{J/\psi}-m_{\eta_c} & = & 28\mbox{\err{2}{2}}\  \mev\hspace{1in}\wilson
\label{eq:hfsw}\\
m_{J/\psi}-m_{\eta_c} & = & 52\mbox{\err{3}{4}}\  \mev\hspace{1in}\clover
\label{eq:hfsc}\eeqn
to be compared to the experimental value of 117(2)~MeV.
The errors quoted in (\ref{eq:hfsw}) and (\ref{eq:hfsc}) are
statistical only, and the reader should bear in mind the uncertainty
in the value of the lattice spacing.
The corresponding values found by El-Khadra et al. \cite{aida} are:
51(3) MeV at $\beta=5.7$, 62(4) MeV at $\beta=5.9$ and 68(5) MeV at
$\beta=6.1$. Extrapolating linearly in $a^2$ to the continuum limit,
these authors quote:
\beq
m_{J/\psi}-m_{\eta_c} = 73\pm 10\,\mev
\eeq
where the error includes an estimate of the systematic uncertainty.
The lattice spacing in this work was determined from the 1P-1S mass
splitting (a quantity which is considerably less sensitive to the
form of the action \cite{aida}).
Thus it appears that there is a difference of about 20 MeV due to
the different action used in ref.\cite{aida}.

\paragraph{Decay Constants} The (dimensionless) decay constant of the
$J/\psi$-meson is defined by:
\beq
\langle0|\bar c(0)\gamma_\mu c(0)|J/\psi\rangle=\epsilon_\mu
\frac{m_{J/\psi}^2}{f_{J/\psi}}
\eeq
where $\epsilon_\mu$ is the polarisation vector of the $J/\psi$.
The measured value of the decay constant is $1/f_{J/\psi} = 0.124(5)$.
In our computations we take for the lattice vector current,
the local operator $Z_V^W\bar c(0)\gamma_\mu c(0)$
when using the Wilson action, and
the ``improved" current
\beq
Z_V^C\,\bar c(x)(1+\frac{ra}{2}\gamma\cdot\dl)\gamma_\mu
(1-\frac{ra}{2}\gamma\cdot\dr)c(x)
\eeq
when using the clover action.
$Z_V^W$ and $Z_V^C$ are the renormalisation constants
(which ensure that the currents are correctly normalised), and can
be evaluated in perturbation theory.
We obtain the results
\beqn
\frac{1}{Z_V^W}\frac{1}{f_{J/\psi}} & = & 0.152\mbox{\err{5}{5}}
\hspace{1.5in}\wilson\label{eq:fjpsiw}\\
\frac{1}{Z_V^C}\frac{1}{f_{J/\psi}} & = & 0.179\mbox{\err{8}{7}}
\hspace{1.5in}\clover\label{eq:fjpsic}
\eeqn
The $Z_V$'s have been calculated to one-loop order, $Z_V^W$=0.83
and $Z_V^C$=0.90 if the lattice bare coupling constant $g_0^2$
is used as the expansion parameter of perturbation theory,
whereas $Z_V^W$=0.71 and $Z_V^C$=0.83 if an ``effective" coupling
$g_{eff}^2=1.75g_0^2$ is used (following suggestions in ref.\cite{lm}).
For example using the effective coupling we find
$1/f_{J/\psi} = 0.108\mbox{\err{4}{4}}$ using the Wilson action,
and $1/f_{J/\psi} = 0.149\mbox{\err{7}{6}}$ with the clover action.
Hence the value of $1/f_{J/\psi}$ is about 30--40\% larger with the
clover action than the Wilson action
(although the uncertainty in the values of the renormalisation
constants should be borne in mind).

The decay constant of the $\eta_c$ has not been measured, however
we present the lattice results in order to compare the values obtained
with the two actions.
The decay constant is defined by
\beq
|\langle0|\bar c(0)\gamma_\mu\gamma_5c(0)|\eta_c(p)\rangle|\equiv
f_{\eta_c}p_\mu
\eeq
(with such a normalisation $f_\pi\simeq 132$ MeV).
For the lattice axial current we take the same operators as in the
vector case with $\gamma_\mu\rightarrow\gamma_\mu\gamma_5$.
The corresponding results are:
\beqn
\frac{1}{Z_A^W}f_{\eta_c}a & = & 0.130\mbox{\err{6}{5}}
\hspace{1.5in}\wilson\label{eq:fetacw}\\
\frac{1}{Z_A^C}f_{\eta_c}a & = & 0.149\mbox{\err{9}{6}}
\hspace{1.5in}\clover\label{eq:fetacc}
\eeqn
$Z_A^W$ and $Z_A^C$ are equal to 0.87 (0.78) and 0.98 (0.97)
when the bare (effective) coupling is used as the expansion parameter.
Thus the values of $f_{\eta_c}$ are also about 30--40\% larger
when determined using the clover action than those obtained
with the Wilson action.

{}From eqs.(\ref{eq:fjpsiw})-(\ref{eq:fetacc}) we note that both
the quantities $1/Z_V\,f_{J/\psi}^{-1}$ and $1/Z_A\,f_{\eta_c}$ are
larger when obtained using the clover action than the Wilson
action.
This is opposite to the results for the corresponding
quantities for the light mesons $\pi$ and $\rho$ \cite{ukqcd1}.
Thus the difference obtained for the decay constants with the two
actions is amplified significantly in the ratios
$\frac{f_{J/\psi}^{-1}}{f_\rho^{-1}}$ and
$\frac{f_{\eta_c}}{f_\pi}$.

Using the chiral extrapolations for $f_\pi$ and $f_\rho^{-1}$, we find
\beqn
\frac{f_{J/\psi}^{-1}}{f_\rho^{-1}} & = & 0.32\mbox{\err{1}{2}} \hspace{1.5in}
\wilson\label{eq:fjpsiwr}\\
\frac{f_{J/\psi}^{-1}}{f_\rho^{-1}} & = & 0.48\mbox{\err{3}{3}}
\hspace{1.5in}\clover\label{eq:fjpsicr}
\eeqn
where the physical value of this ratio is 0.44(2), and
\beqn
\frac{f_{\eta_c}}{f_\pi} & = & 2.3\mbox{\err{5}{3}} \hspace{1.5in}\wilson
\label{eq:fetacwr}\\
\frac{f_{\eta_c}}{f_\pi} & = & 4.0\mbox{\err{12}{9}}
\hspace{1.5in}\clover\label{eq:fetaccr}
\eeqn

In these ratios the dependence on the renormalisation constants
cancels.  The differences between the results for the two actions in
eqs.(\ref{eq:fjpsiwr})-(\ref{eq:fetaccr}) indicate significant errors
due to the finiteness of the lattice spacing (at least, presumably,
for the Wilson action) in the decay constants for the charmonium
system.

\paragraph{Conclusions} In this letter we have shown that the value of
the hyperfine splitting in charmonium in lattice simulations is very
sensitive to the fermion action which is used, and in particular to
the magnetic moment term in the improved action.  Using the same gluon
configurations, we have found that at $\beta=6.2$, the ratio of the
splittings for the clover and Wilson actions is about 1.8 (see
eq.(\ref{eq:ratio}) ).  Even using the clover action, the value we
obtain for the hyperfine splitting is only about one half of the
physical value.  The clover action is a ``tree-level improved
action'', i.e. there are no errors of $O(a)$, but the leading
remaining errors due to the finiteness of the lattice spacing are of
$O(\alpha_s a)$.  Presumably at least some of the discrepancy between
the value in eq.(\ref{eq:hfsc}) and the physical one of 117(2)~MeV is
due to these remaining $O(\alpha_s a)$ errors.  El-Khadra et al. have
tried to reduce these by performing a mean field calculation to
estimate the effects of the higher-order perturbative terms on the
magnetic moment term in the action~\cite{aida,ehkm}.  Their result for
the hyperfine splitting, of about 73~MeV, although larger than that in
eq.(\ref{eq:hfsc}), is still considerably smaller than the physical
value.  In view of the sensitivity of the splitting to the magnetic
moment term in the action, it is likely that at least part of the
discrepancy is still due to the finiteness of the lattice spacing.
Unfortunately it is not possible at present to determine how much of
the discrepency is due to the inadequacy of the mean field
calculation, and how much to other systematic errors (such as
quenching).  We are forced to accept that the hyperfine splitting is
currently not amenable to an accurate lattice determination.
Nevertheless it is reassuring to find that one of the few quantities
for which lattice computations with the Wilson fermion action give a
result which disagrees significantly with experiment, is unusually
sensitive to known systematic errors.

For the decay constants of charmonium we also found significant
differences between the values obtained using the two actions.  For
the ratio $f_{J/\psi}^{-1}/f_\rho^{-1}$ (see eqs.(\ref{eq:fjpsiwr})
and (\ref{eq:fjpsicr}) ) we find a result which is about 75\% of the
physical one with the Wilson action (consistent with the results in
ref.\cite{sixpt4}), and a result which is consistent with the physical
one with the clover action.  For the decay constant of the $\eta_c$
(see eqs.(\ref{eq:fetacwr}) and (\ref{eq:fetaccr}) ) we find a
larger result with the clover than with the Wilson action
(note however that the error for the clover action is large,
this error is dominated by the uncertainty in the extrapolated
value of $f_\pi$).

It will be very interesting to repeat this study for heavy-light
mesons. In particular simulations with Wilson fermions indicate
a substantial violation of the scaling law $f_P\sqrt{M_P}\simeq$
constant (up to logarithmic corrections) \cite{sixpt4,tall}
for the decay constants of heavy-light pseudoscalar mesons $P$.
This would imply a value of $f_B$ of about 200 MeV (one which
is consistent with the simulations using the static
approximation \cite{bmes,wupp}), larger than many earlier
expectations. It is
now important to check whether these results will be stable under
the reduction of the errors of $O(a)$ achieved by the use of the
clover action.

\paragraph{Acknowledgements}

This research is supported by the UK Science and Engineering
Research Council under grant GR/G~32779, by the University of
Edinburgh and by Meiko Limited.  We are grateful to Edinburgh
University Computing Service for use of its DAP 608 for some of
the analysis and, in particular, to Mike Brown for his tireless
efforts in maintaining service on the Meiko i860 Computing
Surface.
CTS acknowledges the support of the UK Science and
Engineering Research Council through the award of a Senior Fellowship.


\begin{thebibliography}{999}

\bibitem{hqs} M.~Bochicchio et al., \np{B372} (1992) 403

\bibitem{sixpt4} A.~Abada et al., \np{B376} (1992) 172
\bibitem{sw} B.~Sheikholeslami and R.~Wohlert, \np{B259} (1985) 572

\bibitem{ukqcd1} The UKQCD Collaboration: C.R.~Allton et al.,
\pl{B284} (1992) 377

\bibitem{light_hadrons} The UKQCD Collaboration: C.R.~Allton et
al., $O(a)$-Improvement for Quenched Light Hadrons at
$\beta=6.2$: Mass Spectrum and Decay Constants, in preparation.

\bibitem{aida} A.X.~El-Khadra, ``Charmonium with Improved Wilson
Fermions II: the Spectrum", Fermilab Preprint FERMILAB-CONF-92/10-T
(1992) (to be published in the proceedings of Lattice 1991)

\bibitem{ehkm} A.X.~El-Khadra, G.~Hockney, A.S.~Kronfeld and
P.B.~Mackenzie,
``A Determination of the Strong Coupling Constant from the Charmonium
Spectrum",
Fermilab Preprint FERMILAB-PUB-91/354-T (1991)

\bibitem{APE}
The APE Collaboaration: S.~Cabasino et al., \pl{B258} (1991) 195.

\bibitem{lm} G.P.~Lepage and P.B.~Mackenzie, Nucl.Phys.
\underline{B(Proc.Suppl.)20} (1991) 173

\bibitem{tall} C.R.Allton et al., \np{B(Proc.Suppl.)20} (1991) 504

\bibitem{bmes} C.R.Allton et al., \np{B349} (1991) 598

\bibitem{wupp} C.Alexandrou et al., \pl{B256} (1991) 60

\end{thebibliography}
\end{document}